# Understanding Hot Carrier Reliability in FinFET Technology from Trap-based Approach

Runsheng Wang[1*], Zixuan Sun[1], Yue-Yang Liu[2], Zhuoqing Yu[1], Zirui Wang[1], Xiangwei Jiang[2], Ru Huang[1]
[1]Institute of Microelectronics, Peking University, Beijing 100871, China ([*]email: r.wang@pku.edu.cn)
[2]State Key Laboratory of Superlattices and Microstructures, Institute of Semiconductors, Chinese Academy of Sciences, Beijing 100083, China

***Abstract*—**In this paper, the recent advances of our studies on hot carrier degradation (HCD) are presented from trap-based approach. The microscopic speculation of interface trap generation is carried out by time-dependent DFT (TDDFT) simulation in "real-time". Two types of oxide traps contributing to HCD are identified from experiments. Combining the contributions of interface and oxide traps, a unified compact model has been proposed which can accurately predict hot carrier degradation and variation in full $V_{gs}/V_{ds}$ bias. The trap locations, degradation contributions and temperature dependence are studied in detail. In addition, the mixed mode reliability of HCD-BTI coupling through self-heating and under off-state stress are discussed.

## I. Introduction

Hot carrier degradation (HCD) is a major concern for device and circuit reliability [1-3]. With the evolution of technology node, self-heating effect (SHE) became a promoter of HCD in FinFET [4, 5], as shown in Fig. 1. However, understanding of physical mechanisms and accurate compact modeling for HCD face great challenges.

On one hand, HCD is highly related to the carrier-induced vibrational excitation of Si-H bond at interface [6-8]. As shown in Fig. 2, Si-H bond dissociation can be triggered by "hot" carrier in single-vibrational excitation (SVE) mode and/or by "cold" carrier in multi-vibrational excitation (MVE) mode [9,10]. Although MVE and SVE are believed to share the physical mechanism of resonance scattering excitation [11], the dynamic process behind the electron/Si-H bond interaction has not been quantified. The above studies are very important to reveal the energy-driven carrier mechanism. On the other hand, since the degradation is ultimately caused by traps, study on the properties of HCD traps provide a new perspective for future research, especially when carrier-based approach become more and more complicated in advanced technology nodes [12].

In this paper, the recent advances of our studies on HCD is presented from trap-based approach. The dynamic process of interface trap generation is demonstrated by time-dependent DFT (TDDFT) simulation for the first time. Then, a trap-based compact model is presented which is unified across different $V_{gs}/V_{ds}$ regions with different carrier-based mechanisms. Furthermore, the properties of different traps (interface trap and two types of oxide traps) are studied in full bias map. Finally, the mixed mode reliability of HCD-BTI coupling is disscused.

## II. Different Traps in HCD

### *A. Interface traps generation: microscopic speculation*

We used the time-dependent density functional theory (TDDFT) method [13] to study the dynamic process of electron injection on the Si-H bond breaking in "real-time". Framework of the TDDFT simulator in this work is shown in Fig. 3.

As shown in Fig. 4, the electron injection will enhance the Si-H bond vibration amplitude, after three extra electrons injected to the Si-H sates. But, Si-H only obtained 0.018eV which is rather small compared to the Si-H bond breaking barrier (i.e. ~3.4 eV for stretching break and ~2.8 eV for bending break [1]). This result suggests a normal Si-H bond at the $Si/SiO_2$ interface can scarcely be broken unless a large number of carriers interact with the bond coherently within the Si-H phonon lifetime, which is only 1~1.5 ns.

We built five possible atomistic structures to verify whether the "Si-HO···H-Si" complex structure could be important to the Si-H bond breaking (Fig. 5). As shown in Fig. 5(c), the structure will be stable in the electron ground state, and at the same time the two H atoms are close to each other (1.75 Å) for possible reaction. The barrier of two H atoms to form a $H_2$ and induce Si-H bond breaking is found to be only 1.17 eV (Fig. 6). As shown in Fig. 7, the energy of the empty Si-H states goes down greatly as the two H atoms approach each other, which tremendously benefits the "cold" electrons interact with the Si-H bond. Furthermore, two different "Si-H···H-O" configurations are demonstrated in Fig. 8. The injected electrons will enhance the attraction force between the two H atoms greatly, and finally cause the breaking of Si-H bond and O-H bond, and the formation of a $H_2$ molecule. A Si dangling bond defect is then generated after this reaction.

These results indicate the essential role of surrounding H atoms in Si-H bond breaking through MVE mode.

### *B. Oxide traps become more important*

$\Delta SS/\Delta V_{th}$ decreases as stress time increases has been observed in Fig. 9(a), suggesting HCD is a combination of interface traps and oxide traps. As the Fig. 9(b) shows the contribution of oxide traps will increase with stress time, which is one of the important mechanisms that cause HCD, especially for pFinFET. SILC spectrums are used to sense the oxide traps as shown in Fig. 9(c), in which two peaks are found for FinFETs (while only one peak for planar HKMG devices [14]). The results indicate that, two types of oxide traps are generated in FinFET under HCD stress. Compared with BTI trap, HCD trap is hardly recoverable. So, the energy gap between occupied state and neutral state of HCD trap is larger than BTI trap. A $V_O$ trap in $SiO_2$ [15] may be a candidate responsible for HCD, while for BTI it has been reported to be induced by the hydroxyl-E' and hydrogen bridge trap [16,17].

## C. *Trap-based unified compact model*

Based on the studies above, we proposed a compact model form trap-based approach in Table I, which is unified across different carrier-based mechanisms (Fig. 10(a)). Note that the SHE should be calibrated at first to get local channel temperature ($T_c$). As shown in Fig. 10 (b), different models can all fit in short term, but they will deviate a lot in long term, suggest that the establishment of an accurate model is essential for the prediction of long-term aging. Our model can accurately predict HCD and its variation in full $V_{gs}/V_{ds}$ bias. The contributions of different traps can be well decomposed (e.g. Fig. 11). More details can be found in [14,18].

## D. *Typical trap locations in FinFET*

The typical lateral locations of traps can be analyzed with forward and reverse mode I-V meaurements [19]. The η' values of different traps are extracted from variation experiments in saturation operation region. As shown in Fig. 12, the η' is larger in the forward mode for interface and oxide trap 1, and larger in the reverse mode for oxide trap 2. It indicates that interface and oxide trap 1 are crowded near the source region, and oxide trap 2 is distributed near the drain side in FinFET.

In order to investigate the typical round-Fin locations of different traps, the η values of different traps are extracted from experiments with our variation model in linear operation region, and compared with the calibrated TCAD results round the Fin [20]. The locations of interface trap and oxide trap 1 are crowded in the middle of Fin side in FinFET, and oxide trap 2 are genrated in the Fin top, as shown in Fig. 13. The different type of oxide traps are genrated in different lattice orientation in FinFET and may share the same origin. For nFinFET, these results suggest that two types of oxide traps are located in the $HfO_2$ layer (or $SiO_2/HfO_2$ interface), different from pFinFET located in the $SiO_2$ layer. The typical locations of different traps in FinFET are summarized in Fig. 14.

# III. Trap Contribution in Full Bias Map

## A. *Temperature dependence*

Activation energy ($E_a$) of different trap has been extracted from experiments as shown in Fig. 15. Since the generation of different traps varies with bias and temperature, HCD does not have a univeral total $E_a$ *(eff)* at different temperature in full bias map [21]. Based on the compact model in Tabel II, we extracted the $E_a$ *(eff)* at different chuck temperatures (Figs. 16 vs. 17). $E_a$*(eff)* of high-$V_{ds}$ and low-$V_{gs}$ region move from negative value to positive value, indicating the oxide trap become more important as the chuck tempreature increases in nFinFET. $E_a$*(eff)* increases with chuck temperature in pFinFET, indicating the interface trap contribution increases at higher temperature in a narrow bias region.

## B. *Time dependence*

As shown in Figs. 18-19, the dominant bias region of different traps can vary with time. In nFinFET, interface trap is dominant in the high-$V_{ds}$ and low-$V_{gs}$ region, while the low-$V_{ds}$ and high-$V_{gs}$ region is dominated by oxide trap 2 in the short time. With the aging time increase, the high bias region is dominated by oxide trap 1. For pFinFET, the interface trap is not dominate in full bias map, and oxide trap 2 is dominant in the high bias region. And the dominant region of oxide trap 1 is expanded with aging time increase.

# IV. Mixed Mode Reliability of HCD-BTI Coupling

## A. *Coupling through self-heating effect*

The contributions of BTI and HCD can be simply split by our model (Table I) in full bias map, but SHE is a tricky factor impacting their contributions. In Fig. 20, we plot the $\Delta V_{th}$ (HCD) / $\Delta V_{th}$ (BTI) in full bias map with SHE. Under DC stress, HCD dominates at high-$V_{ds}$ and low-$V_{gs}$ region because of the $V_{ds}$ reduced BTI and increased carrier energy. Under 1MHz AC HCD stress, SHE is close to DC condition, the territory of HCD enlarges due to the BTI fast traps recovery during the low-level voltage period. When the frequency reaches 1GHz, the influence of the SHE can be ignored and thus the local channel temperature is low. In nFinFET, the territory of PBTI expands in the high-$V_{gs}$ and low-$V_{ds}$ region, because oxide trap 2 with a greater $E_a$ than PBTI trap. In pFinFET, NBTI trap has a larger $E_a$ than HCD trap, so the proportion of NBTI narrows as the frequency increases to 1GHz.

## B. *Coupling under off-state stress*

Off-state degradation is a mixed mode of HCD-BTI in nanoscale FinFET [22]. We have found new off-state degradation mechanisms (Fig. 21). It is different from traditional understanding of hole trapping from band-to-band tunneling at the gate-drain overlap under off-state stress. As device scaling, more off-state carriers and stronger lateral field lead to intense impact ionization in the channel, making the off-state degradation become severer (Fig. 22). A compact model has been proposed, which can accutely predict off-state reliability and separate the contributions of HCI and BTI components.

# V. Conclusion

The recent advances of our studies on HCD are summarized, and providing a trap-based perspective for future research. The properties of different traps are discussed, and a unified compact model is presented.

Acknowledgment

This work was supported by NSFC (61874005, 61927901), and the 111 Project (B18001).

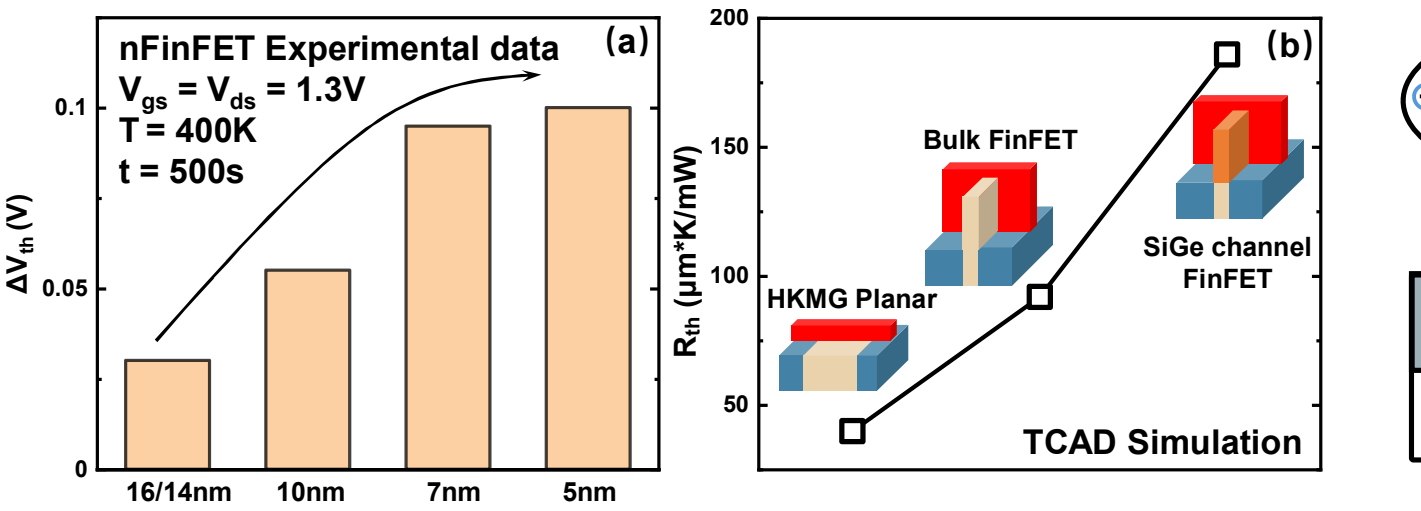

**Fig. 1.** (a) With the scaling of CMOS devices, HCD became more important in advanced FinFET. (b) 3D structure and low thermal conductivity materials induced severe self-heating effect will enhance HCD.

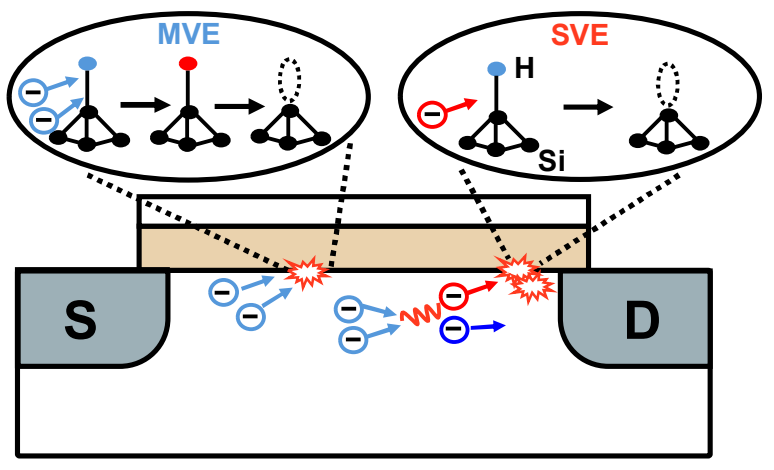

**Fig. 2.** A schematic of Si-H bond dissociation mechanisms with the single-vibrational excitation and the multi-vibrational excitation.

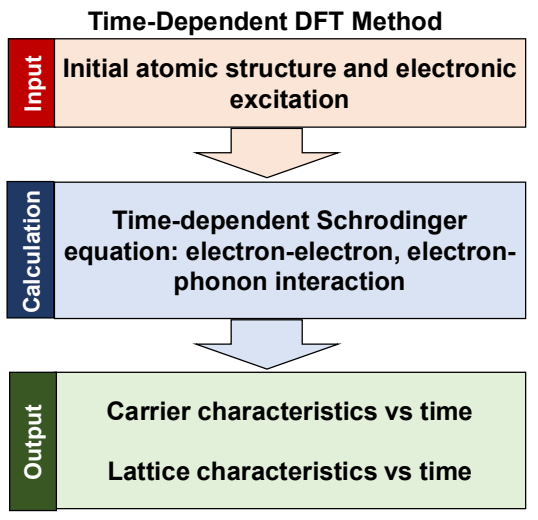

**Fig. 3.** Framework of TDDFET simulator in this work.

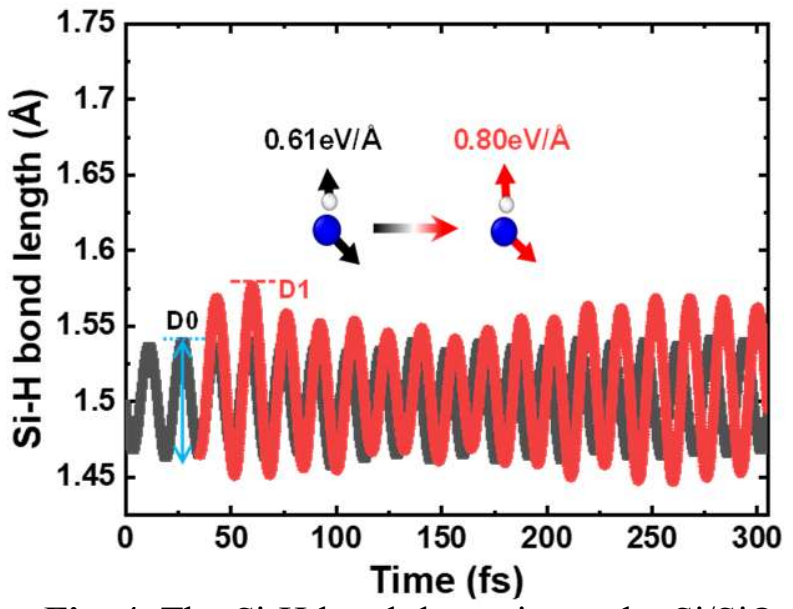

**Fig. 4.** The Si-H bond dynamics at the $Si/SiO_2$ interface. Electrons are injected to the Si-H states at 35fs. The electron injection will enhance the Si-H bond vibration amplitude, but the energy gained by Si-H is only 0.018eV.

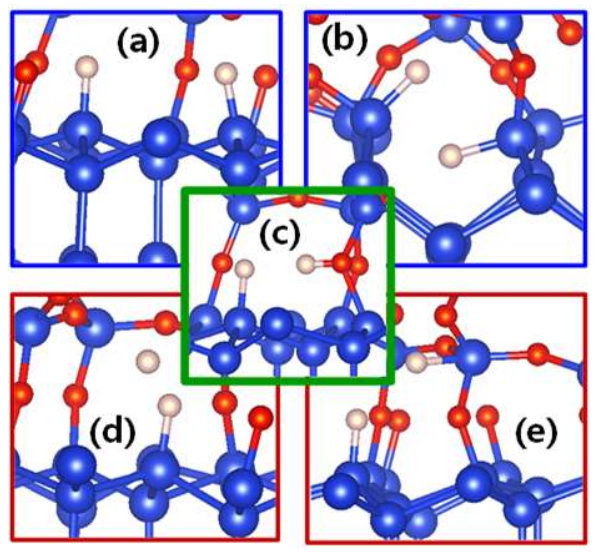

**Fig. 5.** Five possible structures are constructed. The distance between two H atoms are (a) 3.30 Å, (b) 2.06 Å, (c) 1.75 Å. (d-e) Si-H bond are unstable in the ground state.

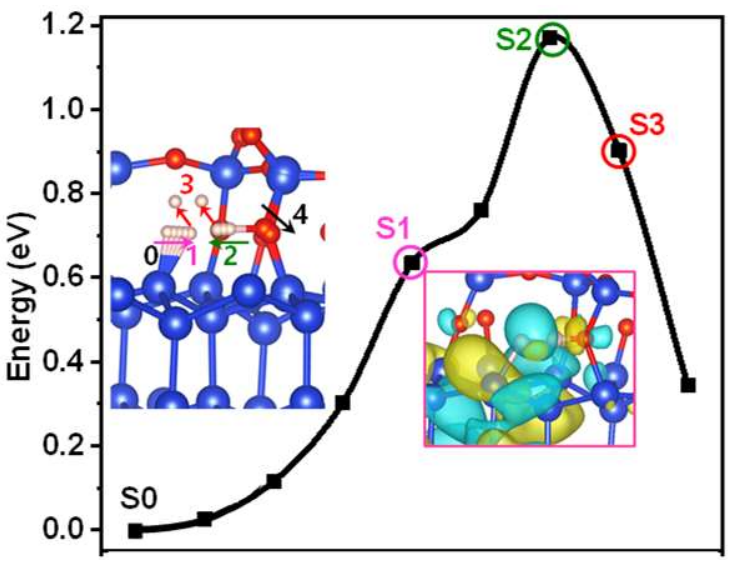

**Fig. 6.** The reaction barrier of the precursor with "Si-H···H-O" structure in Fig. 5(c).

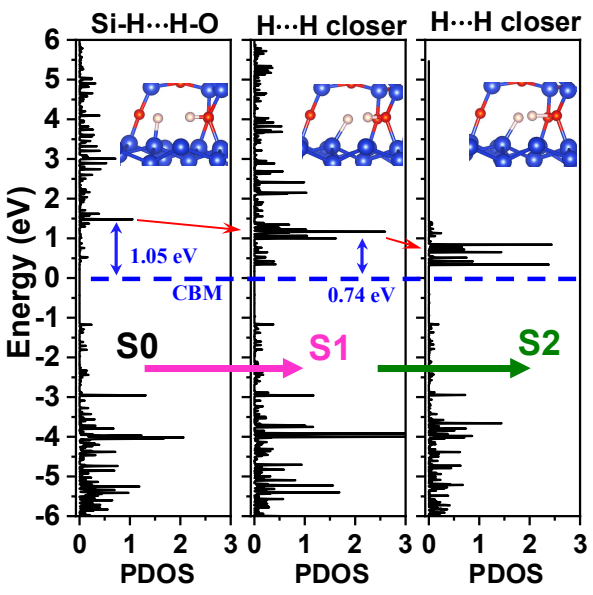

**Fig. 7.** Energy decrease of the Si-H states when the two H atoms approach each other.

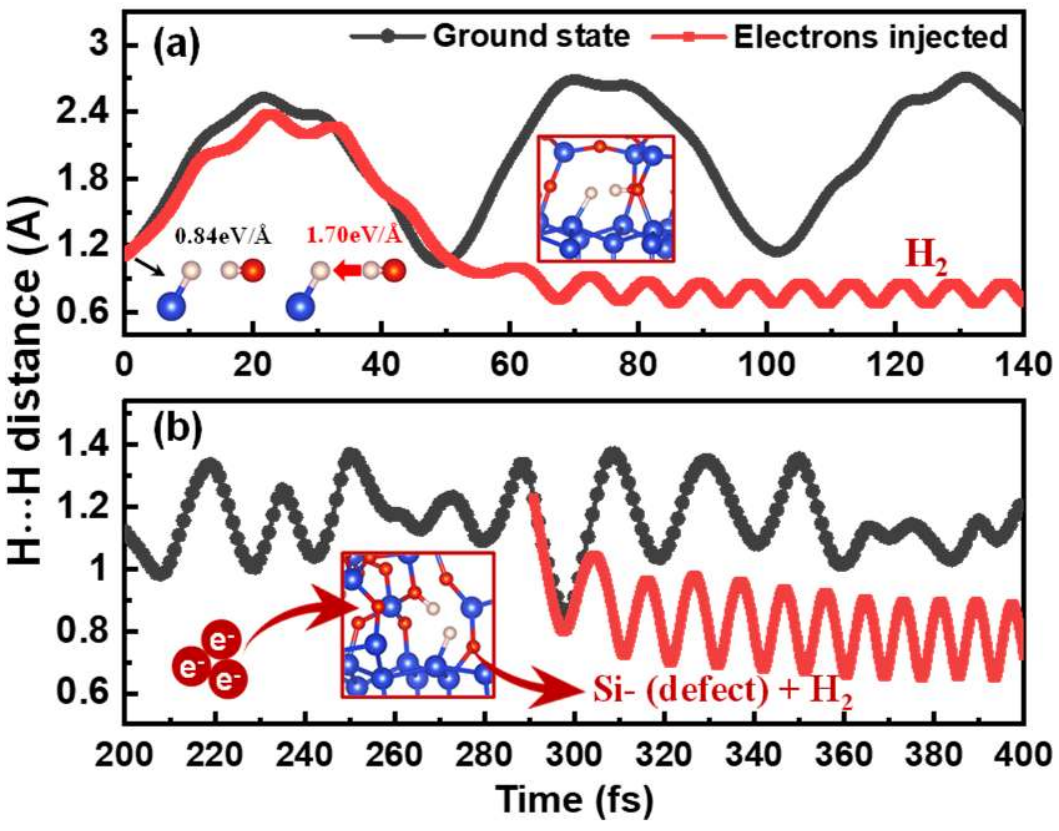

**Fig. 8.** The carrier induced Si-H bond breaking and $H_2$ formation in two different "Si-H···H-O" configurations. The additional H atom bonds with an O atom (a) at the exact $Si-SiO_2$ interface and (b) inside the amorphous oxide.

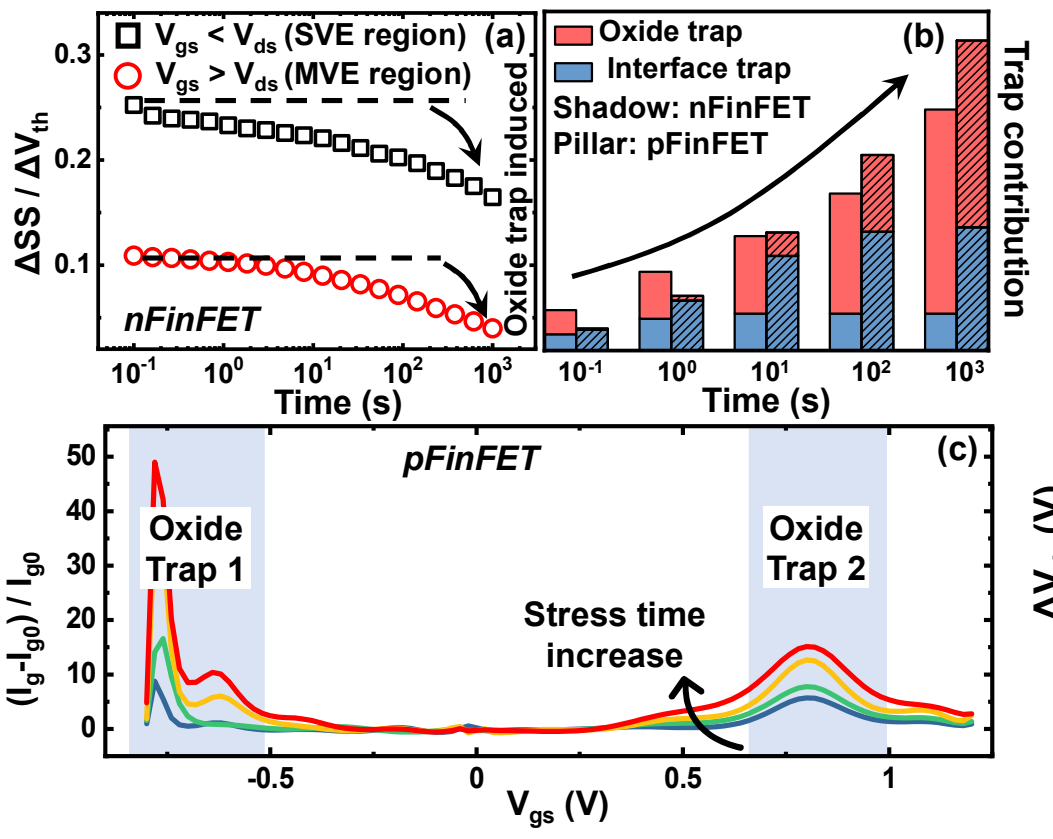

**Fig. 9.** (a) Oxide trap induced $\Delta SS/\Delta V_{th}$ decrease with aging time increase. (b) Oxide traps are more important over time. (c) SILC spectrums found two peaks, suggesting there are two types of oxide trap in FinFET.

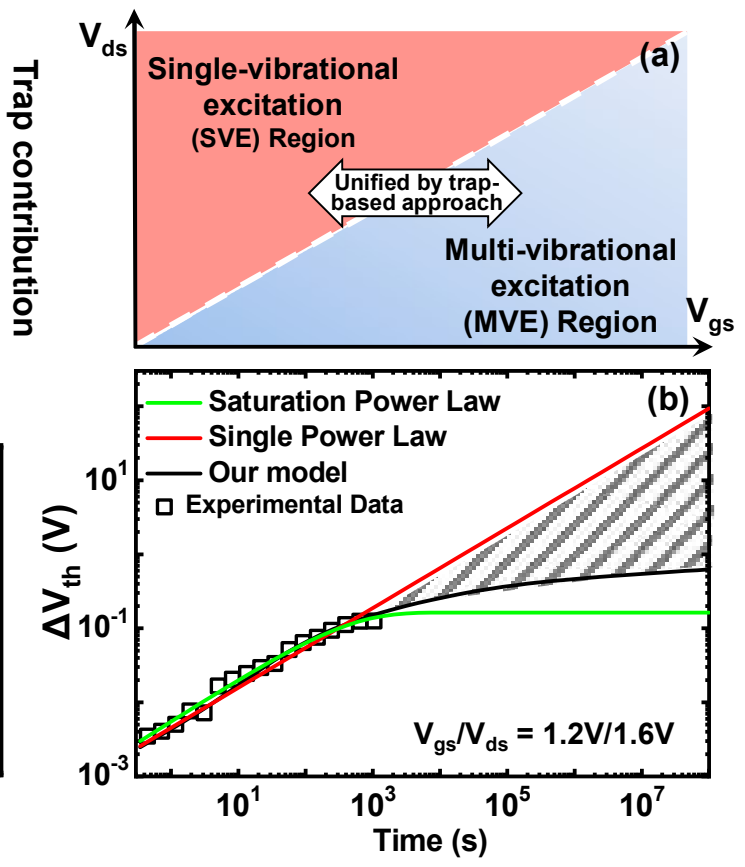

**Fig. 10.** (a) Our model is unified in different mechanism region during HCD. (b) Different models can fit short-term experimental data well, but they will deviate a lot in long-term.

| Trap-based Compact Model for HCD in FinFET | | | |
|---|---|---|---|
| Nominal model | | Variability model | |
| Total Degradation | $TD = HCD_{Interface} + HCD_{Oxide\,1} + HCD_{Oxide\,2} + k \times BTI$ | Total Average Degradation | $\mu_{Total} = \mu_{Interface} + \mu_{Oxide\,1} + \mu_{oxide2}$ |
| BTI Ratio | $k = 1 - 0.5 \times V_{ds}/V_{gs}$ | Total variation of Degradation | $\sigma^2_{Total} = \sigma^2_{Interface} + \sigma^2_{Oxide\,1} + \sigma^2_{Oxide\,2}$ |
| Interface Trap | $HCD_{Interface} = N_0 \times [1 - \exp(-AR_i \times t^n)]$ | | |
| Oxide Trap 1 | $HCD_{Oxide\,1} = AR_1 \times \log(1 + c_1 t)$ | Interface Trap | $\sigma^2_{Interface} = 2\eta_1 \times \mu_{Interface}$ |
| Oxide Trap 2 | $HCD_{Oxide\,2} = AR_2 \times \log(1 + c_2 t)$ | Oxide Trap 1 | $\sigma^2_{oxide1} = 2\eta_2 \times \mu_{oxide\,1}$ |
| Age Rate (AR) | $A(V_{gs} - V_{th})^m \exp\left(\frac{-b}{V_{ds} - V_{dsat}}\right) \exp\left(\frac{-E_a}{K_B T_c}\right)$ | Oxide Trap 2 | $\sigma^2_{oxide2} = 2\eta_3 \times \mu_{oxide\,2}$ |

**Table 1.** The trap-based compact models are proposed, including nominal and variation model.

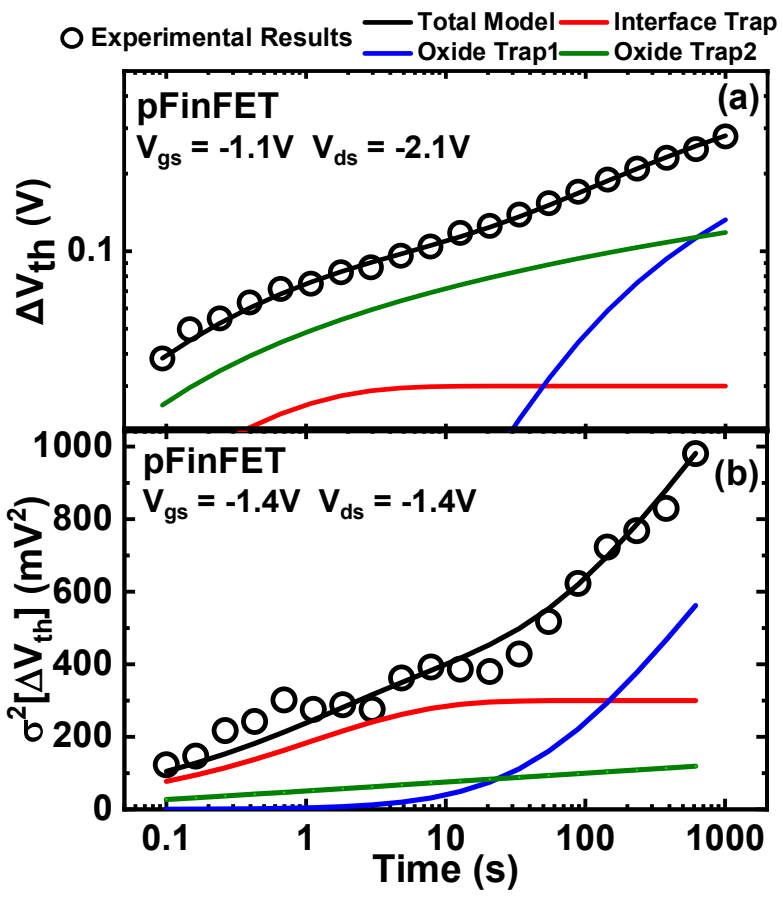

**Fig. 11.** Our compact model can well-decompose the (a) degradation and (b) variation into the contribution of different types of traps.

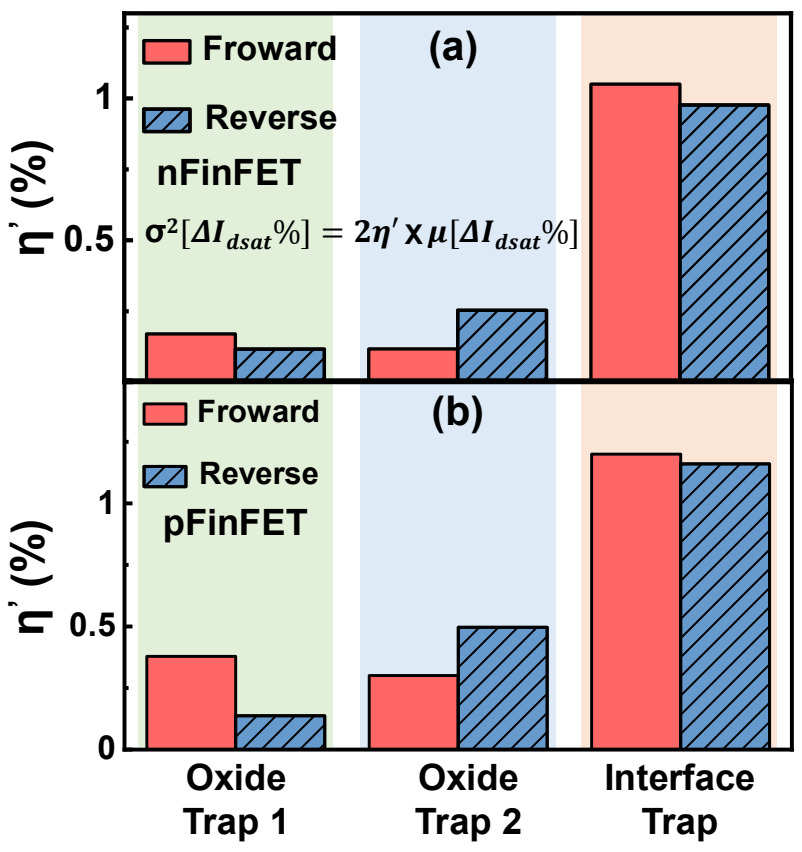

**Fig. 12.** Experimental results of the single-trap-induced average degradation of interface trap and oxide traps 1 and 2 in both forward and reverse modes.

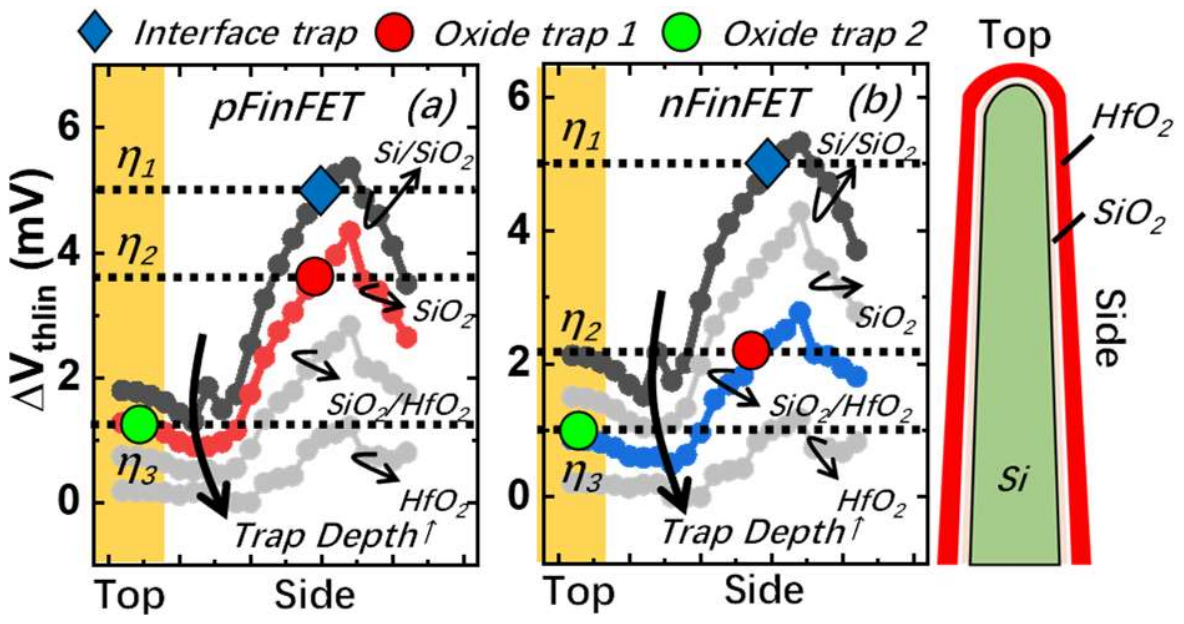

**Fig. 13**. Calibrated TCAD simulations of the single-trap-induced average $\Delta V_{th}$ at different round-Fin locations in (a) pFinFETs and (b) nFinFETs. Symbols represent the typical "average" round-Fin locations of different traps.

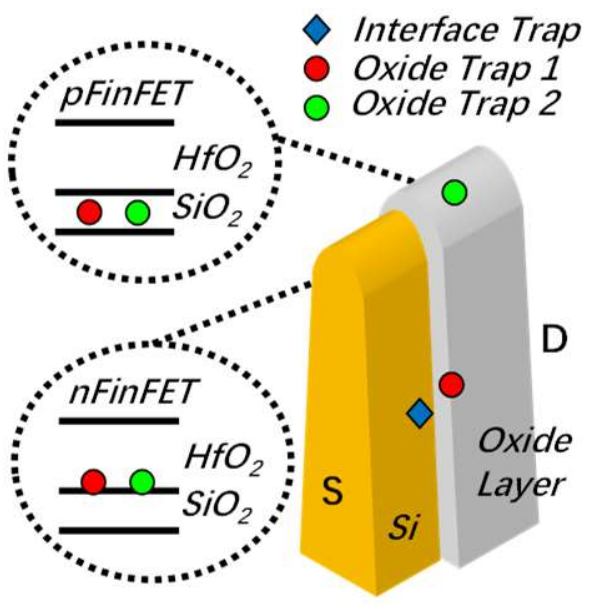

**Fig. 14**. Typical locations of the interface traps and two types of oxide traps generated by HCD in both n- and p-type FinFETs.

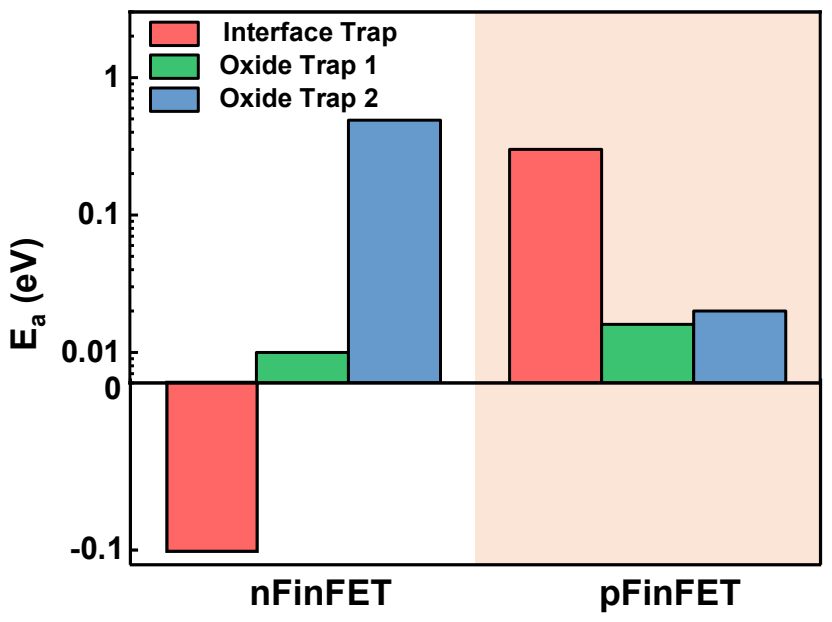

**Fig. 15**. The extracted $E_a$ of interface trap, oxide trap1 and oxide trap 2 for nFinFET and pFinFET during HCD stress.

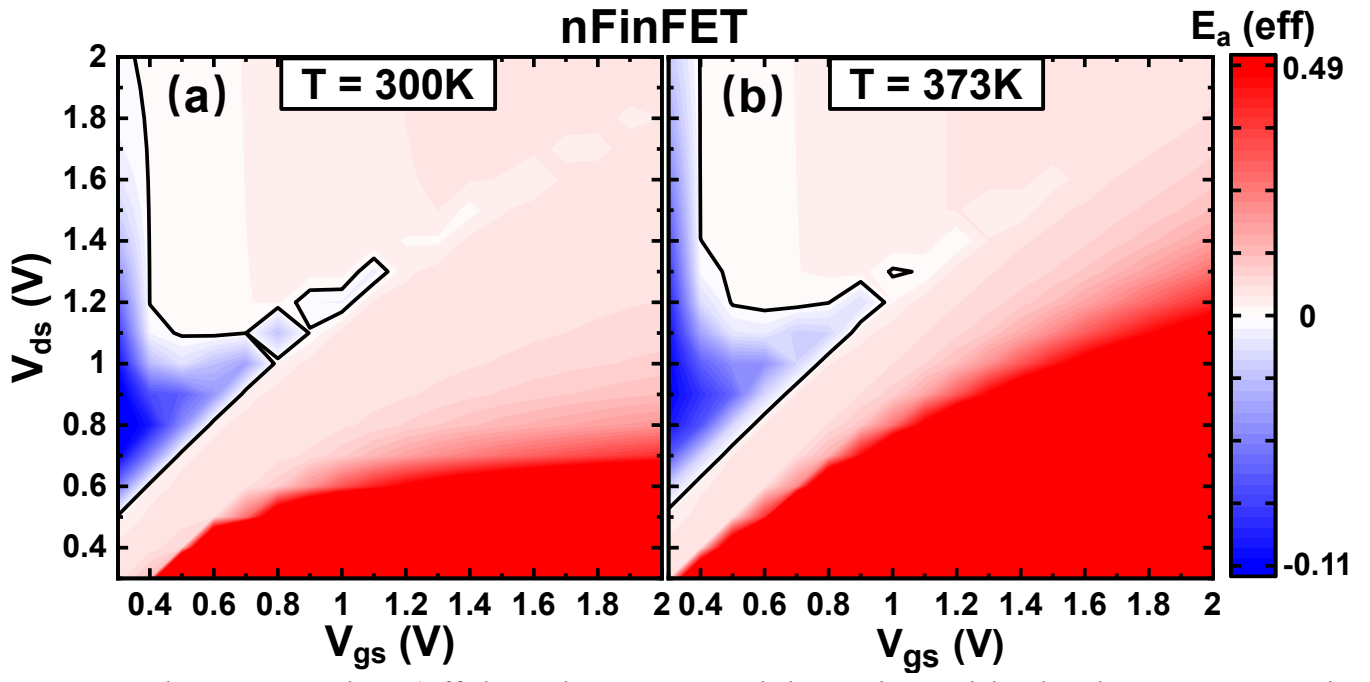

**Fig. 16**. The extracted $E_a(eff)$ based on our model varying with chuck temperature in nFinFET.

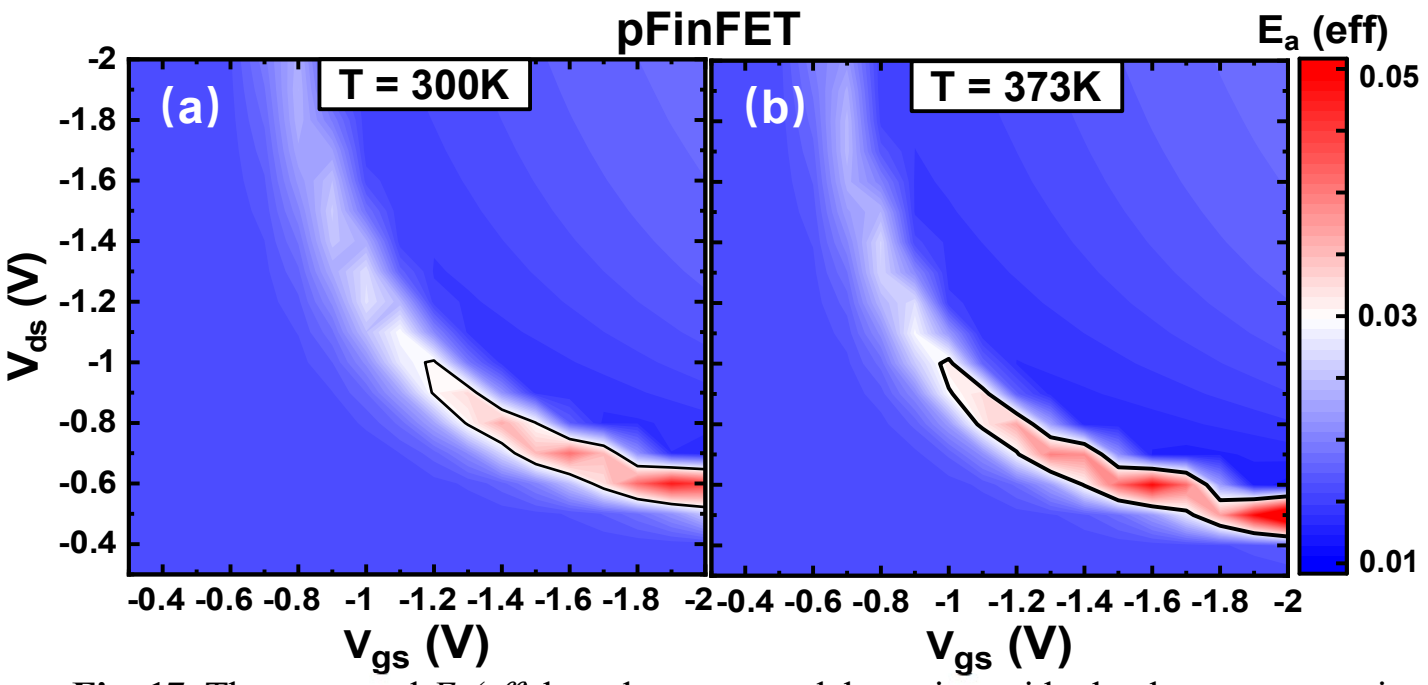

**Fig. 17**. The extracted $E_a(eff)$ based on our model varying with chuck temperature in pFinFET.

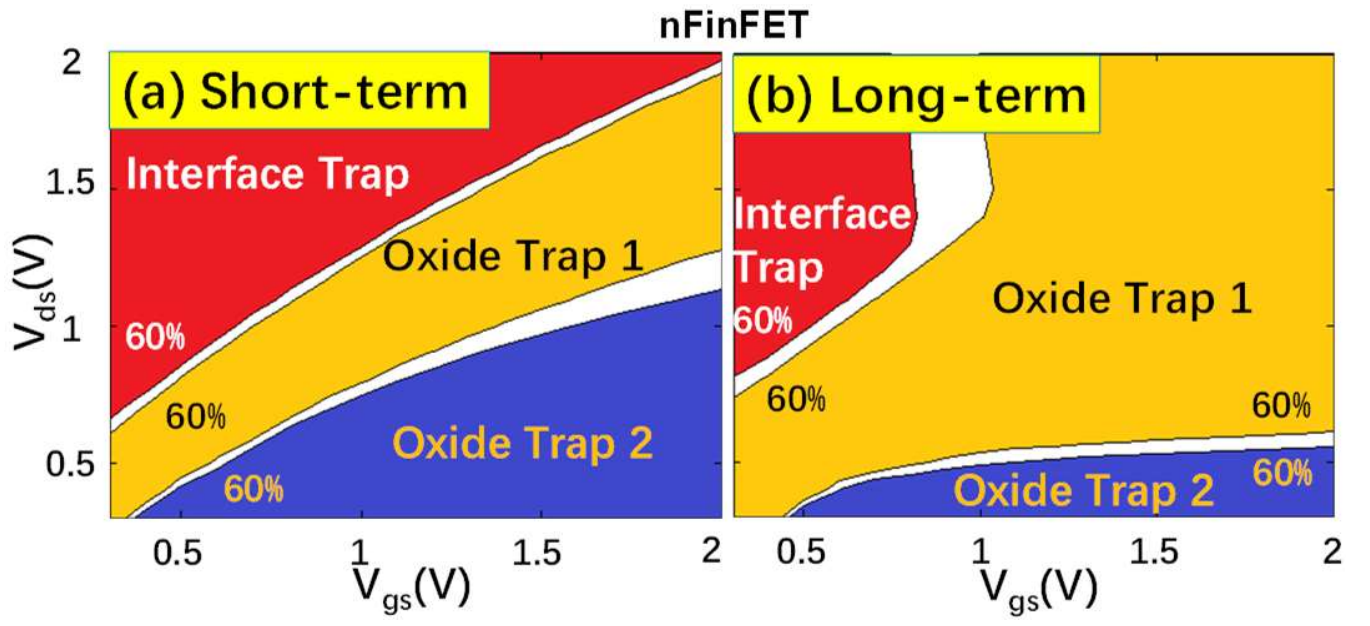

**Fig. 18**. Normalized results of the "territories" of different traps with the aging time in nFinFET. Interface trap governs the high-$V_{ds}$ and low-$V_{gs}$ corner and oxide trap 2 governs the low-$V_{ds}$ and high-$V_{gs}$ corner (T=300K).

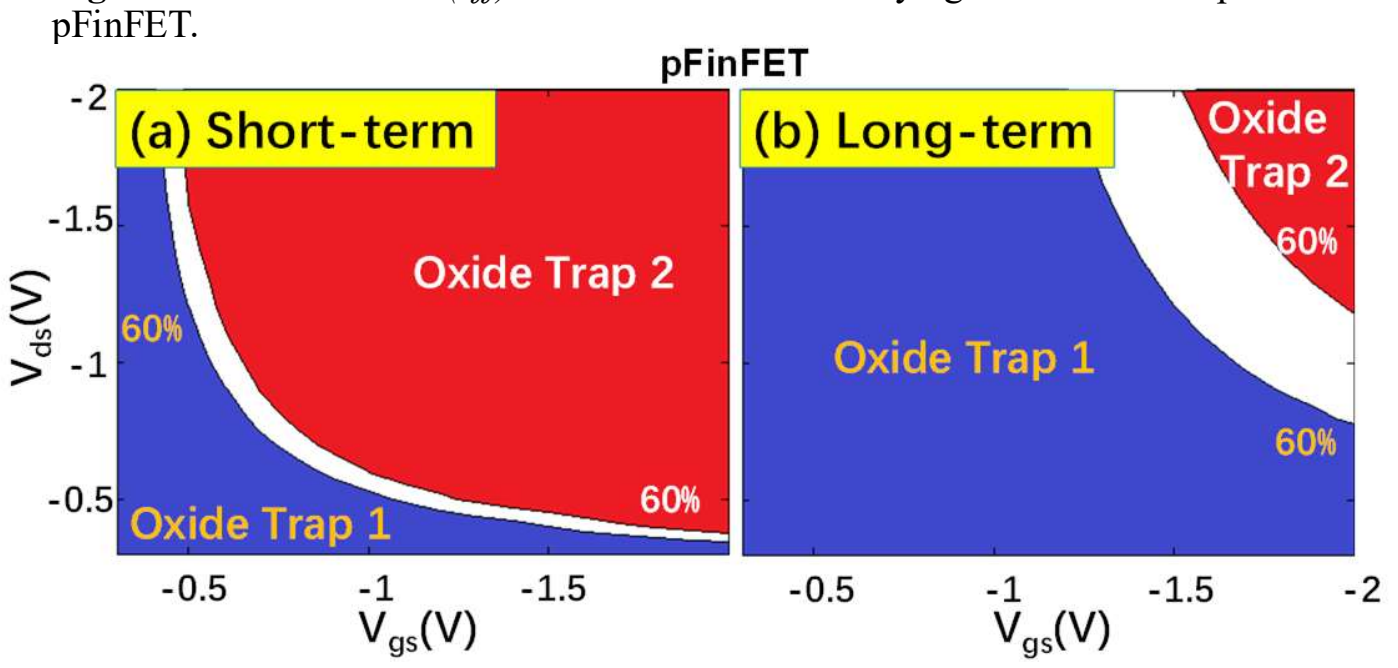

**Fig. 19**. pFinFET normalized results of the "territories" of different traps. Oxide trap 1 will be more important in long-term degradation (T=300K).

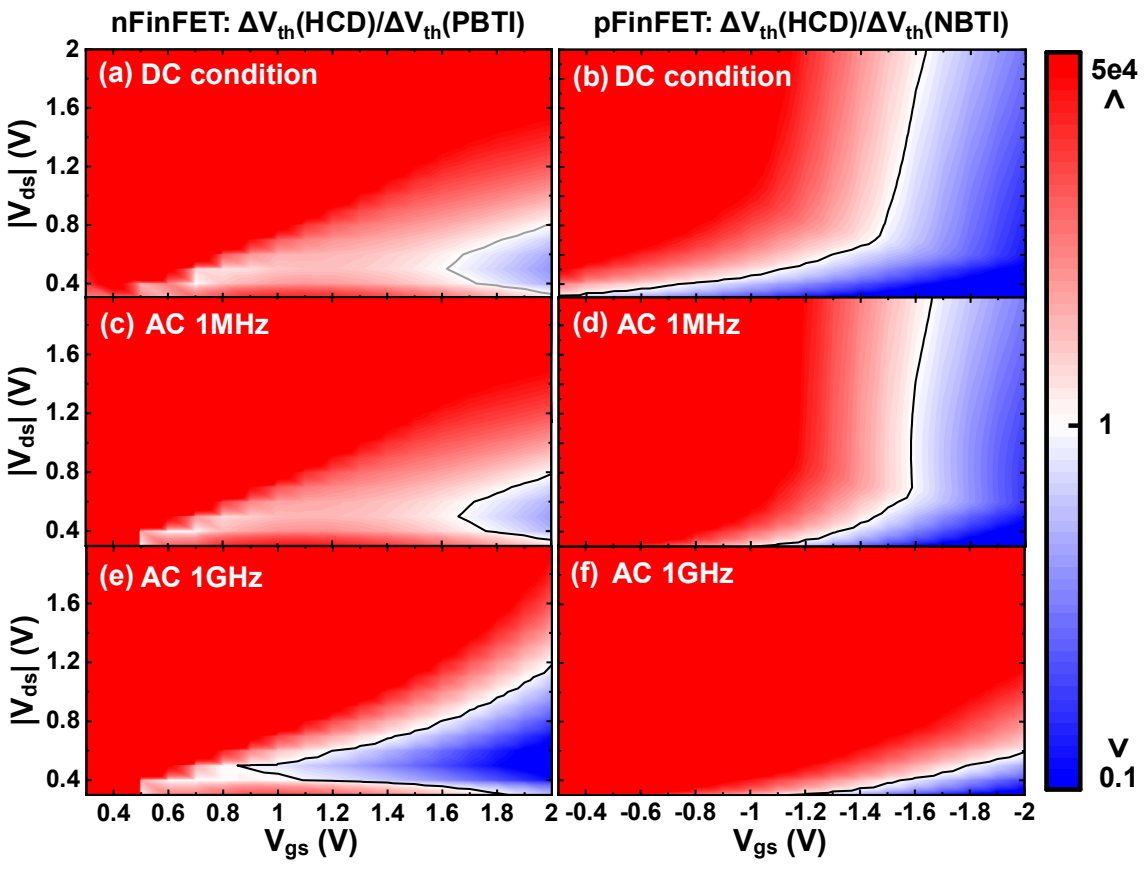

**Fig. 20**. The ratio of HCD and BTI contributions in full map with self-heating effect (T=300K). DC stress condition: (a) nFinFET, (b) pFinFET. 1MHz AC HCD stress ($V_{gs}$ is AC signal with duty factor DF=0.5, $V_{ds}$ is DC bias): (c) nFinFET, (d) pFinFET. 1GHz AC HCD stress: (e) nFinFET, (f) pFinFET.

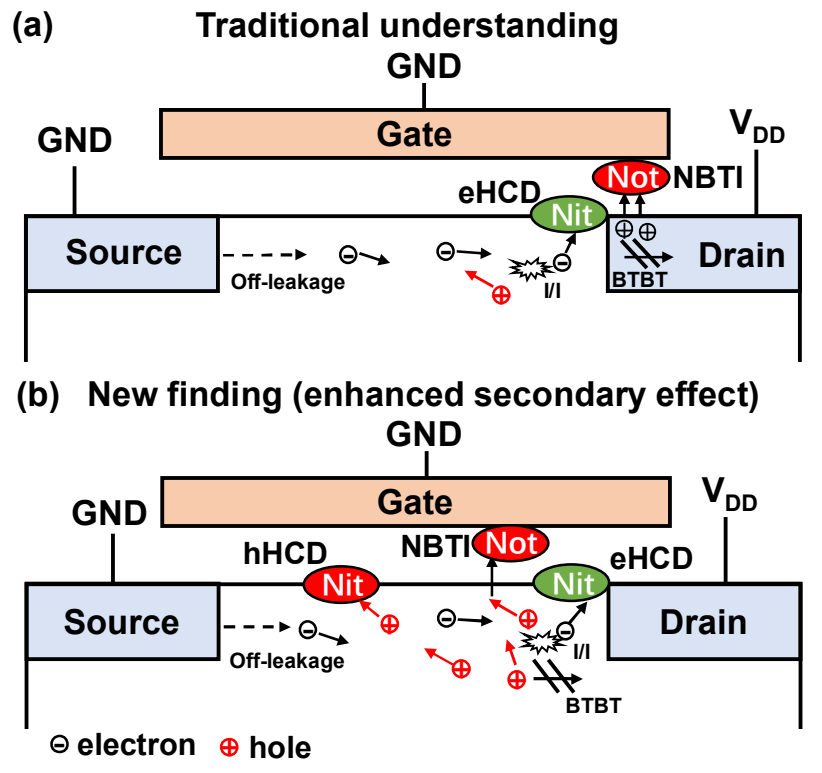

**Fig. 21** Off-state degradation in NMOS, (a) traditional understanding: oxide trapping at gate-drain overlap and electrons I/I near the drain region. (b) New finding: the secondary holes come from electrons I/I and BTBT in channel-drain region. Those secondary holes can generate NBTI and hHCD traps.

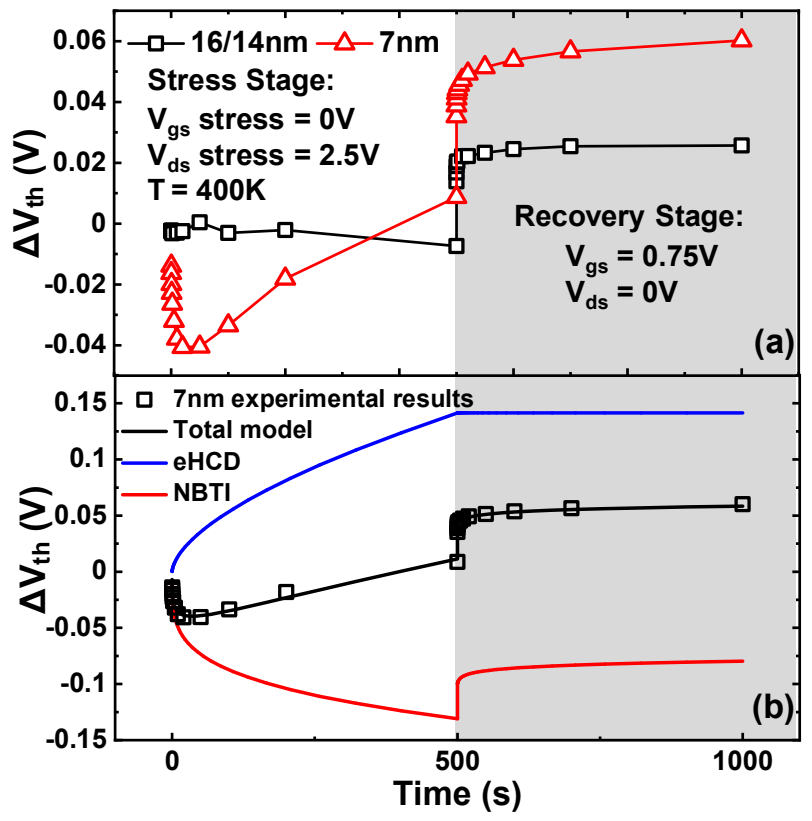

**Fig. 22**. (a) Comparison of $V_{th}$ degradation between 16/14 nm and 7 nm nFinFET under off-state stress. (b) The model can well-decompose the total degradation into the contribution of different type traps.